\documentclass[a4paper]{jpconf}
\usepackage{graphicx}
\usepackage{amssymb}
\usepackage{color}

\newcommand{\beq}{\begin{equation}}
\newcommand{\eeq}{\end{equation}}
\newcommand{\barr}{\begin{eqnarray}}
\newcommand{\earr}{\end{eqnarray}}

\newcommand{\andy}[1]{ }
\def\cH{{\mathcal{H}}}
\def\cV{{\mathcal{V}}}
\def\cU{{\mathcal{U}}}
\def\cP{{\mathcal{P}}}

\def\As{{\mathcal{A}}}
\def\bra#1{\langle #1 |}
\def\ket#1{| #1 \rangle}

\renewcommand{\Re}{\mathrm{\,Re\,}}

\begin{document}
\title{Quantum Zeno dynamics and quantum Zeno subspaces}

\author{Paolo Facchi$^1$, Giuseppe Marmo$^2$, Saverio Pascazio$^3$ }

\address{$^1$ Dipartimento di Matematica, Universit\`a di Bari  \\
and Istituto Nazionale di Fisica Nucleare, Sezione di Bari, I-70125
Bari, Italy}
\address{$^2$
Dipartimento di Scienze Fisiche, Universit\`a di Napoli ``Federico
II" \\
and Istituto Nazionale di Fisica Nucleare, Sezione di Napoli,
I-80126 Napoli, Italy}
\address{$^3$
Dipartimento di Fisica, Universit\`a di Bari
\\
and Istituto Nazionale di Fisica Nucleare, Sezione di Bari, I-70126
Bari, Italy }

\ead{saverio.pascazio@ba.infn.it}

\begin{abstract}
A quantum Zeno dynamics can be obtained by means of
frequent measurements, frequent unitary kicks or a strong continuous
coupling and yields a partition of the total Hilbert space into
quantum Zeno subspaces, among which any transition is hindered. We
focus on the ``continuous" version of the quantum Zeno effect and
look at several interesting examples. We first analyze these
examples in practical terms, towards applications, then propose a
novel experiment.
\end{abstract}

%\pacs{PACS numbers: 03.65.Xp}

\section{Introduction }
 \label{sec-introd}
 \andy{intro}

The quantum Zeno effect is intertwined with the name of George
Sudarshan, who cast the problem in a rigorous mathematical
framework and proposed the classical allusion to the sophist
philosopher in a seminal article written in collaboration with
Baidyanaith Misra in 1977
\cite{Misra}.  Somewhat curiously, the most remarkable practical
application of the quantum Zeno effect consists in reducing (and
eventually suppressing) decoherence and dissipation, which are other
issues in which George Sudarshan was a protagonist, by establishing
a firm physical and mathematical description that is known nowadays
as the Gorini-Kossakowski-Sudarshan-Lindblad equation
\cite{GKSL}. The aim of this article is to illustrate the main
features of the quantum Zeno effect (QZE), clarify what is a quantum
Zeno dynamics, discuss examples and recent experiments and propose
applications.

The quantum Zeno effect has an interesting history. It was first
understood by von Neumann \cite{von}, who proved that any given
quantum state $\phi$ can be ``steered" into any other state $\psi$, by applying a suitable series of
measurements. If $\phi$ and $\psi$ coincide (modulo a phase factor),
the evolution yields, in modern language, a quantum Zeno effect.
After 35 years (!) Beskow and Nilsson \cite{Beskow} considered a
particle in a bubble chamber (thought of as an apparatus that
``continuously checks" whether the particle has decayed) and
wondered whether this ``measurement" mechanism can hinder decay.
This interesting intuition was then reconsidered by other authors,
both from a physical \cite{Khalfin68,peresetal} and more genuinely
mathematical perspective \cite{Friedman72}. Notice that a rigorous
formulation hinges upon difficult mathematical issues
\cite{ExnerIchinose}, most of which are yet unsolved \cite{exnerreview}.

It was not until 1988 that Cook realized that the quantum Zeno
\emph{effect} (not a ``paradox" as people tended to regard it)
could be \emph{tested} on oscillating (two- or three-level) systems
\cite{Cook}. This proposal deviated from the original framework, that dealt
with \textit{bona fide} unstable systems
\cite{Beskow,Misra}, but was nonetheless an interesting and
concrete idea, that led to a beautiful experimental test a few
years later, performed by Itano and collaborators
\cite{Itano}.
The debate that followed \cite{Itanodisc} motivated novel
experimental tests, involving diverse physical systems, such as
photon polarization, chiral molecules and ions \cite{kwiat} and very
recently Bose-Einstein condensates \cite{ketterle} and cavity QED
\cite{parisqed}. New experiments have been proposed with neutron
spin \cite{VESTA}.

The presence of a short-time quadratic (Zeno) region for a
\emph{bona fide} unstable quantum mechanical system (particle
tunnelling out of a confining potential) was experimentally
confirmed by Raizen and collaborators in 1997 \cite{Wilkinson}. A
few years later, the same group demonstrated the Zeno effect
(hindered evolution by frequent measurements)
\cite{raizenlatest}. This experiment also proved the
occurrence of the curious inverse (or anti) Zeno effect (IZE)
\cite{antiZeno,heraclitus}, first suggested in
1983~(!), according to which the evolution can be
\emph{accelerated} if the measurements are frequent, but not
\emph{too} frequent. The IZE has remarkable links with chaos
and dynamical localization \cite{varieidee}.

The QZE arises as a straightforward consequence of general features
of the Schr\"odinger equation, that yield quadratic behavior of the
survival probability at short times \cite{strev,zenoreview}. It is
usually understood as the hindrance of the dynamics out of the
intial state due to frequent von Neumann measurements. Nowadays, in
view of possible applications, this picture appears too restrictive
in two respects. First, the QZE does not necessarily freeze the
dynamics. On the contrary, for frequent projections onto a
\emph{multi}dimensional subspace, the system can evolve away from
its initial state, although it remains in the ``Zeno subspace"
defined by the measurement \cite{compactregularize}. The resulting
constrained evolution is called ``quantum Zeno dynamics"
\cite{theorem}.
Second, the QZE can be reformulated in terms of a ``continuous coupling"
 \cite{zenoreview}, without making use of projection
operators and non-unitary dynamics, obtaining the same physical
effects. We emphasize that the idea of a continuous formulation
of the QZE is not new \cite{varicont,peresetal}, but has appeared in
the literature in different contexts and at different times. These
two observations will enable us to focus on possible interesting
applications and novel experimental proposals.

This article is organized as follows. We start by reviewing some
notions related to the (familiar) ``pulsed" formulation of the Zeno
effect in Sec.\ \ref{sec-dpw}. We then summarize the celebrated
Misra and Sudarshan theorem in Sec.\
\ref{sec-msth}. This theorem is generalized in Sec.\
\ref{sec-partial}, in order to
accommodate nonselective measurements. We briefly introduce the
``kicked" version of the QZE in Sec.\ \ref{sec-qmaps} and the
continuous formulation in Sec.\
\ref{sec-contQZE}.
The three procedures (pulsed, kicked and continuous QZE) are
compared in Sec.\
\ref{sec-comm}, where their physical equivalence is discussed.
We then look at several elementary but relevant examples in Secs.\
\ref{sec-3lev}-\ref{sec-cook}. This is a central part of the
article, where a new experiment is proposed. We conclude with a few
comments in Sec.\ \ref{sec-concl}.

\section{Preliminaries: the ``pulsed" formulation}
\label{sec-dpw}
\andy{sec-dpw}

We start off by giving an elementary introduction to the QZE, in its
most familiar formulation, in terms of  von
Neumann measurements represented by one-dimensional projections. Let $H$ be the total Hamiltonian of a quantum
system Q and $\ket{\psi_0}$ its initial state at $t=0$. The survival
amplitude and probability in state $\ket{\psi_0}$ at time time $t$
read ($\hbar=1$)
\andy{uno}
\barr
\As (t) &=& \langle \psi_0|e^{-iHt}|\psi_0\rangle , \label{eq:unoa}\\
p(t) &=& |\As (t)|^2 =|\langle \psi_0|e^{-iHt}|\psi_0\rangle |^2
\label{eq:unob}
\earr
and a short-time expansion yields a quadratic behavior
\andy{quadratic}
\barr
p(t) &\sim & 1 - t^2/\tau_{\mathrm{Z}}^2, \\
\tau_{\mathrm{Z}}^{-2} & \equiv & \langle \psi_0|H^2|\psi_0\rangle - \langle
\psi_0|H|\psi_0\rangle^2 ,
\label{eq:quadratic}
\earr
where $\tau_{\mathrm{Z}}$ is the Zeno time \cite{hydrogen}.
Observe that if the Hamiltonian is divided into a free and an
(off-diagonal) interaction parts
\andy{Hdiv}
\barr
& & H = H_0 + H_{\mathrm{int}},
\nonumber \\
& & H_0\ket{\psi_0} =
\omega_0\ket{\psi_0}, \nonumber \\
& & \bra{\psi_0}H_{\mathrm{int}}\ket{\psi_0}=0, \label{eq:Hdiv}
\earr
the Zeno time reads
\andy{tzoff}
\beq
\tau_{\mathrm{Z}}^{-2} = \bra{\psi_0}H_{\mathrm{int}}^2\ket{\psi_0}
\label{eq:tzoff}
\eeq
and depends only on the interaction Hamiltonian.

\begin{figure}[t]
\begin{center}
\includegraphics[width=8cm]{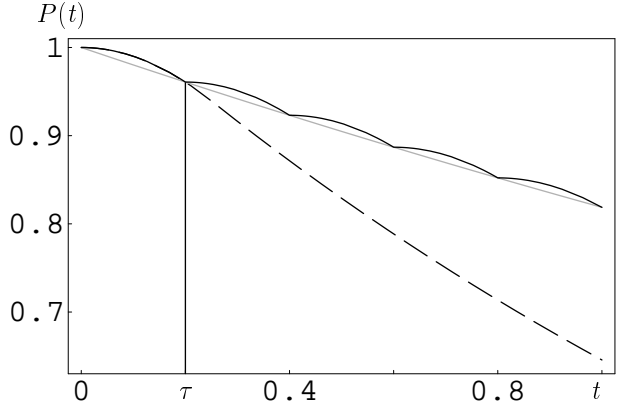}
\end{center}
\caption{Quantum Zeno effect in the ``pulsed" formulation.
The dashed (full) line is the survival probability without (with)
measurements. The gray line is the interpolating exponential
(\ref{eq:survN0}). The units on the abscissae are arbitrarily
chosen for illustrative purposes.}
\label{fig:zenoevol}
\end{figure}

Let us now show how frequent projective measurements can hinder
evolution away from the initial state. Perform $N$ measurements at
time intervals $\tau=t/N$, in order to check whether the system is
still in its initial state $\ket{\psi_0}$. The survival probability
$p^{(N)}(t)$ at time $t$ reads
\andy{survN}
\beq
p^{(N)}(t)=p(\tau)^N = p\left(t/N\right)^N
\sim\exp\left(-t^2/\tau_{\mathrm{Z}}^2 N\right) \stackrel{N
\rightarrow\infty}{\longrightarrow} 1 .
\eeq
For large $N$ the evolution is slowed down and in the $N\to\infty$
limit the evolution is completely hindered. Notice that the
survival probability after $N$ pulsed measurements ($t=N\tau$) is
interpolated by an exponential law
\cite{heraclitus}
\andy{survN0}
\beq
p^{(N)}(t)=p(\tau)^N=\exp(N\log p(\tau))=
\exp(-\gamma_{\mathrm{eff}}(\tau) t) ,
\label{eq:survN0}
\eeq
with an effective decay rate
\andy{eq:gammaeffdef}
\beq
\gamma_{\mathrm{eff}}(\tau) \equiv -\frac{1}{\tau}\log p(\tau) =
-\frac{2}{\tau}\log |\As (\tau)| =-\frac{2}{\tau}\Re [\log
\As(\tau)] \ge0 \; . \label{eq:gammaeffdef}
\eeq
For $\tau\to 0 $ ($N \to \infty$) one gets  $p(\tau) \sim
\exp (-\tau^2/\tau_{\mathrm{Z}}^2)$, whence
\beq
\gamma_{\mathrm{eff}}(\tau)\sim \tau/\tau_{\mathrm{Z}}^2,
\qquad \tau\to 0.
\label{eq:lingammaeff}
\eeq
The Zeno evolution for ``pulsed" measurements is pictorially
represented in Figure
\ref{fig:zenoevol}.

\section{Misra and Sudarshan's theorem}
\label{sec-msth}
\andy{msth}

The formulation of the previous section is very intuitive and does
not rest on a firm mathematical ground. The existence of the moments
of the Hamiltonian and the convergence of the expansions are all
taken for granted and subtle field-theoretical issues related to the
Fermi golden rule \cite{Fermi} are not considered
\cite{QFT,hydrogen}.

The formulation by Misra and Sudarshan consists in a theorem that
requires a few preliminary notions. We first introduce incomplete
measurements, represented by multidimensional projections, by
applying the von Neumann-L\"{u}ders \cite{von,Luders} formulation in
terms of projection operators and adopting some definitions given by
Schwinger \cite{Schwinger59} (but see also Peres
\cite{Peres98}). We will say that a measurement is ``incomplete"
if some outcomes are lumped together. This happens, for example, if
the experimental equipment has insufficient resolution (and in this
sense the information on the measured observable is incomplete).
The projection operator $P$, that selects a particular lump, is
therefore multidimensional.

Let ${\cal H}$ be the Hilbert space of system Q and let the evolution be
described by the unitary operator $U(t)=\exp(-iHt)$, where $H$ is a
time-independent lower-bounded Hamiltonian. Let $P$ be a projection
operator and $\textrm{Ran}P=\cH_P$ its range, with $\mathrm{dim}
\cH_P = \mathrm{Tr} P=s \leq \infty$ (not necessarily finite dimensional). We assume that the
initial density matrix $\rho_0$ of system Q belongs to ${\cal H}_P$:
\andy{inprep}
\beq
\rho_0 = P \rho_0 P , \qquad \mathrm{Tr} [ \rho_0 P ] = 1 .
\label{eq:inprep}
\eeq
Under the sole action of the Hamiltonian $H$ (no measurements),
the state at time $t$ reads
\andy{noproie}
\beq
\rho (t) = U(t) \rho_0 U^\dagger (t)
  \label{eq:noproie}
\eeq
and the survival probability (namely the probability that the
system is still in ${\cal H}_P$) at time $t$ reads
\andy{stillun}
\beq
p(t) = \mathrm{Tr} \left[ U(t) \rho_0 U^\dagger(t) P \right] .
\label{eq:stillun}
\eeq
No distinction is made between one- and multi-dimensional
projections.

The above evolution is ``undisturbed:" Q evolves under the sole
action of its Hamiltonian for a time $t$, without undergoing any
measurement process. We now perform a
\textit{selective} measurement at time $\tau$, in order to check
whether Q has ``survived" inside ${\cal H}_P$. By ``selective", we
mean that we filter out the survived component and stop the other
ones. (Think for instance of decomposing a spin in a Stern-Gerlach
setup and absorbing away the unwanted components.) The state of Q
changes into
\andy{proie}
\beq
\rho_0 \rightarrow \rho(\tau) = \frac{P U(\tau) \rho_0 U^\dagger(\tau)
P}{p(\tau)} ,
\label{eq:proie}
\eeq
where
\andy{probini}
\barr
p(\tau) = \mathrm{Tr} \left[ U(\tau) \rho_0 U^\dagger(\tau) P
\right] = \mathrm{Tr} \left[V(\tau) \rho_0 V^\dagger(\tau)
\right],
  \qquad V(\tau) \equiv P U(\tau)P
\label{eq:probini}
\earr
is the survival probability in ${\cal H}_P$. The QZE is the
following. We prepare Q in the initial state $\rho_0$ at time 0 and
perform a series of (selective) $P$-observations at time intervals
$\tau=t/N$ (by this we mean that Q is found in $\cH_P$ at every
step). The state at time $t$ reads
\andy{Nproie}
\beq
\rho^{(N)}(t) = \frac{V_N(t) \rho_0 V_N^\dagger(t)}{p^{(N)}(t)} ,
\qquad
V_N(t) = \left[ V\left(\frac{t}{N}\right)\right]^N,
\label{eq:Nproie}
\eeq
where
\andy{probNob}
\begin{equation}
p^{(N)}(t) = \mathrm{Tr} \left[ V_N(t) \rho_0 V_N^\dagger(t)
\right]
\label{eq:probNob}
\end{equation}
is the survival probability in ${\cal H}_P$. Equations
(\ref{eq:Nproie})-(\ref{eq:probNob}) are the formal statement of the
QZE, according to which very frequent observations modify the
dynamics of the quantum system: under general conditions, if $N$ is
sufficiently large, transitions outside ${\cal H}_P$ are inhibited.

The $N \rightarrow \infty$ limit requires some technical hypotheses:
assume that the strong limit
\andy{slim}
\beq
\cV (t) \equiv \lim_{N \rightarrow \infty} V_N(t)
  \label{eq:slim}
\eeq
exists  $\forall t>0$. The final state of Q is then
\andy{infproie}
\beq
\rho (t) = \lim_{N\to\infty} \rho^{(N)}(t)=\cV(t) \rho_0
\cV^\dagger (t)
  \label{eq:infproie}
\eeq
and the survival probability in $\cH_P$ is
\andy{probinfob}
\beq
\cP (t) \equiv \lim_{N \rightarrow \infty} p^{(N)}(t)
   = \mathrm{Tr} \left[ \cV(t) \rho_0 \cV^\dagger(t) \right].
\label{eq:probinfob}
\eeq
By assuming the strong continuity of $\cV(t)$ at $t=0$
\andy{phgr}
\beq
\lim_{t \rightarrow 0^+} \cV(t) = P, \label{eq:phgr}
\eeq
Misra and Sudarshan proved that under general conditions the
operators $\cV(t)$ exist $\forall t \in \mathbb{R}$ and form a
semigroup. Moreover, by time-reversal invariance
\andy{VVdag}
\beq
\cV^\dagger (t) = \cV(-t), \label{eq:VVdag}
\eeq
one gets $\cV^\dagger (t) \cV(t) =P$. This implies, by
(\ref{eq:inprep}), that
\andy{probinfu1}
\beq
\cP(t)=\mathrm{Tr}\left[\rho_0 \cV^\dagger(t)\cV(t)\right] =
\mathrm{Tr} \left[ \rho_0 P \right] = 1 . \label{eq:probinfu1}
\eeq
If the particle is very frequently observed, in order to check
whether it has survived inside $\cH_P$, it will never make a
transition to $\cH_P^\perp$. In general, if $N$ is sufficiently
large in (\ref{eq:Nproie})-(\ref{eq:probNob}), transitions outside
${\cal H}_P$ are inhibited. However, if $N$ is not
\emph{too} large the system can display an inverse Zeno effect
\cite{antiZeno,heraclitus}, by which decay is
accelerated. Both effects have been demonstrated
\cite{raizenlatest}. We will not elaborate on this here.

A few comments are in order. Notice that the dynamics
(\ref{eq:Nproie})-(\ref{eq:probNob}) is not reversible. On the other
hand, the dynamics in the $N \to
\infty$ limit is often time reversible \cite{compactregularize}
(although, in general, the operators $\cV(t)$ form a
\textit{semi}group). Observe also that the Misra and Sudarshan
theorem \textit{does not} state that the system
\textit{remains} in its initial state, after the series of very
frequent measurements. Rather, the system \textit{evolves} in the
subspace $\cH_P$, instead of evolving ``undisturbed" in the total
Hilbert space $\cH$. The limiting \emph{Zeno dynamics} within
$\cH_P$ is governed by (\ref{eq:slim}), which can be a \emph{unitary
group}. Therefore, starting from the dynamics (\ref{eq:Nproie}),
which is irreversible and probability-nonconserving, one may end up
with a fully unitary evolution. Unitarity is recovered in the
limit. The features of this evolution in some simple cases will be
the object study of the following sections. We anticipate that if
$\cH_P\subset D(H)$, the domain of the Hamiltonian $H$, the limiting
time evolution has the explicit form
\beq
\cV (t)= P \exp(-i P H P t), \label{eq:cVfin1}
\label{eq:boundlimit}
\eeq
namely $\cV(t)$ is unitary within $\cH_P$ and is generated by the
self-adjoint Hamiltonian $PHP$: reversibility is recovered in the $N
\to \infty$ limit. In particular, Eq.\ (\ref{eq:boundlimit}) is always valid for bounded Hamiltonian, since $D(H)=\cH$.
In more general cases (infinite dimensional projectors,
$s=\mathrm{dim} \cH_P=\infty$, unbounded $H$) one can always
formally write the limiting evolution in the form (\ref{eq:cVfin1}),
but has to define the meaning of $P H P$. In such a case one has to
study the self-adjointness of the formal limiting Hamiltonian $PHP$
\cite{Friedman72,compactregularize,ExnerIchinose,exnerreview}.

\section{Extension of the Misra and Sudarshan theorem: The quantum Zeno subspaces}
 \label{sec-partial}
 \andy{partial}

We now consider more general measurements and extend the Misra and
Sudarshan theorem in order to accommodate multiple projectors. We
will say that a measurement is ``nonselective"
\cite{Schwinger59} if the measuring apparatus does not ``select"
the different outcomes, so that all the ``branch waves" undergo the
whole Zeno dynamics. In other words, a nonselective measurement
destroys the phase correlations between different branch waves,
provoking the transition from a pure state to a mixture. Let
\beq
\{P_n\}_n, \qquad
P_nP_m=\delta_{mn}P_n,\qquad  \sum_n P_n=1 ,
\eeq
be a (countable) collection of projection operators and
$\mathrm{Ran}P_n=\cH_{P_n}$ the relative subspaces. This induces a
partition on the total Hilbert space
\beq
\label{eq:partition}
\cH=\bigoplus_n \cH_{P_n}.
\eeq
Consider the associated nonselective measurement described by the
superoperator \cite{von,Luders}
\beq
\label{eq:superP} \hat P \rho=\sum_n P_n \rho P_n.
\eeq
The free evolution reads
\beq
\hat U_t \rho_0=U(t) \rho_0 U^\dagger(t),\qquad U(t)=\exp(-i H t)
\eeq
and the Zeno evolution after $N$ measurements in a time $t$ is
governed by the superoperator
\beq
\hat V^{(N)}_t=\hat P\left(\hat U \left(t/N \right)\hat
P\right)^{N-1} .
\eeq
This yields the evolution
\beq
\rho(t)=\hat V^{(N)}_t \rho_0 =\sum_{n_1,\dots,n_N}V_{n_1\dots
n_N}^{(N)}(t)\; \rho_0\; V_{n_1\dots n_N}^{(N)\dagger}(t) ,
\eeq
where
\barr
V_{n_1\dots n_N}^{(N)}(t)  = P_{n_N} U\left(t/N\right) P_{n_{N-1}}
\cdots P_{n_2} U\left(t/N\right) P_{n_1}, \label{eq:boo}
\earr
which should be compared to Eq.\ (\ref{eq:Nproie}). We follow
Misra and Sudarshan \cite{Misra} and assume, as in Sec.\
\ref{sec-msth}, the existence of
the strong limits ($t\in\mathbb{R}$)
\andy{slims}
\beq
\cV_n (t)=\lim_{N\to\infty} V_{n\dots n}^{(N)}(t) , \qquad \lim_{t
\rightarrow 0} \cV_n(t) = P_n , \quad \forall n \ .
\label{eq:slims}
\eeq
Then $\cV_n(t)$  form a semigroup for every $n$ and
\beq
\cV_n^\dagger(t)\cV_n(t)=P_n.
\eeq
Moreover, one can show that
\andy{fulldiag}
\beq
\lim_{N\to\infty} V_{n\dots n'\dots}^{(N)}(t) = 0, \qquad
\mathrm{for}\quad n'\neq n ,
\label{eq:fulldiag}
\eeq
strongly. Notice that, for any \emph{finite} $N$, the off-diagonal operators
(\ref{eq:boo}) are in general nonvanishing, i.e.\ $V_{n\dots
n'\dots}^{(N)}(t) \neq 0$ for $n'\neq n$. It is only in the limit
(\ref{eq:fulldiag}) that these operators vanish. This is
because $U\left(t/N\right)$ provokes transitions among different
subspaces $\cH_{P_n}$. By Eqs.\
(\ref{eq:slims})-(\ref{eq:fulldiag}) the final state is
\andy{rhoZ}
\barr
\rho(t)=\hat \cV_t\rho_0 =\sum_n \cV_n(t) \rho_0 \cV_n^\dagger(t),
\quad \mathrm{with} \quad \sum_n \cV_n^\dagger(t)\cV_n(t)=\sum_n
P_n=1 .\;\; \label{eq:rhoZ}
\earr
The components $\cV_n(t) \rho_0 \cV_n^\dagger(t)$ make up a block
diagonal matrix: the initial density matrix is reduced to a
mixture and any interference between different subspaces
$\cH_{P_n}$ is destroyed (complete decoherence). In conclusion,
\andy{probinfu}
\beq
p_n(t) =  \mathrm{Tr} \left[\rho(t) P_n\right]=
\mathrm{Tr}\left[\rho_0 P_n\right]=p_n(0) , \quad \forall n .
\label{eq:probinfu}
\eeq
Probability is conserved in each subspace and no probability
``leakage" between any two subspaces is possible: the total Hilbert
space splits into invariant subspaces and the different components
of the wave function (or density matrix) evolve independently within
each sector. One can think of the total Hilbert space as the shell
of a tortoise, each invariant subspace being one of the scales.
Motion among different scales is impossible. (See Fig.\
\ref{fig:tortoise} in the following.)

If $\cH_{P_n}\subset D (H)$,  then the limiting evolution
operator $\cV_n(t)$ (\ref{eq:slims}) within the subspace
$\cH_{P_n}$ has the form (\ref{eq:cVfin1}),
\beq
\cV_n (t)=P_n \exp(-i P_n H P_n t)  \label{eq:cVfin} ,
\eeq
$P_n H P_n$ is bounded and self-adjoint and $\cV_n(t)$ is unitary in $\cH_{P_n}$.
Precautions are necessary in more general cases, as mentioned after
Eq.\ (\ref{eq:cVfin1}).

The result (\ref{eq:probinfu1}) is reobtained when $p_n(0)=1$ for
some $n$, in (\ref{eq:probinfu}): the initial state is then in one
of the invariant subspaces and the survival probability in that
subspace remains unity. In conclusion, when the necessary care is
taken for mathematical rigor and subtleties, the Zeno evolution can
be written
\barr
\rho(t)&=&\hat \cV_t\rho_0 =\sum_n \cV_n(t) \rho_0 \cV_n^\dagger(t),
\\
\cV_n (t)&=& P_n \exp(-i H_{\mathrm{Z}}t )
\earr
where
\begin{equation}
H_{\mathrm{Z}}=\hat P H =\sum_n P_n H P_n
\label{eq:ZenoHam}
\end{equation}
is the ``Zeno" Hamiltonian.

\section{Unitary kicks\index{unitary ``kicks"}}
\label{sec-qmaps}
\andy{sec-qmaps}

The formulation of the preceding section hinges upon projections
\textit{\`{a} la} von Neumann. Projections are instantaneous
processes, yielding the collapse of the wave function (an ultimately
nonunitary process). However, one can obtain the QZE without making
use of nonunitary evolutions. In this section we further elaborate
on this issue, obtaining the QZE by means of a \emph{generic}
sequence of frequent \emph{instantaneous unitary} processes. We will only give the main
results, as additional details and a complete proof, which is
related to von Neumann's ergodic theorem \cite{ReedSimon}, can be
found in \cite{FLP}.

Consider the dynamics of a quantum system Q undergoing $N$ ``kicks"
$U_{\mathrm{kick}}$ (instantaneous unitary transformations) in a
time interval $t$
\andy{eq:BBevol}
\begin{equation}
\label{eq:BBevol}
U_N(t)=\left[U_{\mathrm{kick}}U\left(\frac{t}{N}\right)
\right]^N .
\end{equation}
In the large $N$ limit, since the dominant contribution is
$U_{\mathrm{kick}}^N$, one considers the sequence of unitary
operators
\andy{eq:sequence}
\begin{equation}
\label{eq:sequence}
V_N(t)=U_{\mathrm{kick}}^{\dagger N} U_N(t)=
U_{\mathrm{kick}}^{\dagger N}\left[U_{\mathrm{kick}}
U\left(\frac{t}{N}\right) \right]^N
\end{equation}
and its limit
\andy{eq:limseq}
\begin{equation}
\label{eq:limseq}
\cU(t) \equiv \lim_{N\to\infty} V_N(t).
\end{equation}
One can show that
\andy{eq:eqUz}
\begin{equation}
\label{eq:eqUz}
\cU(t)= \exp(-i H_{\mathrm{Z}} t),
\end{equation}
where
\andy{eq:eqHz}
\begin{equation}
\label{eq:eqHz}
H_{\mathrm{Z}} = \hat P H = \sum_n P_n H P_n
\end{equation}
is the Zeno Hamiltonian, $P_n$ being the spectral projections of
$U_{\mathrm{kick}}$
\andy{eq:specdec}
\begin{equation}
U_{\mathrm{kick}}=\sum_n e^{-i\lambda_n} P_n . \quad
(e^{-i\lambda_n}\neq e^{-i\lambda_l}, \; \mbox{for} \; n\neq l).
\label{eq:specdec}
\end{equation}
In conclusion
\begin{eqnarray}
U_N(t)&\sim& U_{\mathrm{kick}}^N \cU(t)= U_{\mathrm{kick}}^N
\exp(-i H_{\mathrm{Z}} t)\\
&=&\exp\left(-i \sum_n  N \lambda_n P_n + P_n H P_n t \right).
\label{eq:evolNZ}
\end{eqnarray}
The unitary evolution (\ref{eq:BBevol}) yields therefore a Zeno
effect\index{quantum Zeno effect} and a partition of the Hilbert
space into Zeno subspaces\index{quantum Zeno subspaces}, like in the
case of repeated projective measurements discussed in the previous
section. The appearance of the Zeno subspaces is a direct
consequence of the wildly oscillating phases between different
eigenspaces of the kick.

\section{Continuous coupling}
 \label{sec-contQZE}
 \andy{contQZE}

The formulation of the preceding sections hinges upon instantaneous
processes, that can be unitary or nonunitary. However, the main
features of the QZE can be obtained by making use of a continuous
coupling, when the external system takes a sort of steady ``gaze" at
the system of interest. The mathematical formulation of this idea is
contained in a theorem
\cite{theorem} on the (large-$K$) dynamical evolution
governed by a \emph{generic} Hamiltonian of the type
\andy{HKcoup}
\begin{equation}
H_K= H + K H_{\mathrm{c}} ,
 \label{eq:HKcoup}
\end{equation}
which again need not describe a \textit{bona fide} measurement
process. $H$ is the Hamiltonian of the quantum system investigated
and $H_{\mathrm{c}}$ can be viewed as an ``additional" interaction
Hamiltonian performing the ``measurement." $K$ is a coupling
constant.

Consider the time evolution operator
\begin{equation}
U_{K}(t) = \exp(-iH_K t) . \label{eq:measinter}
\end{equation}
In the ``infinitely strong coupling" (``infinitely quick
detector") limit $K\to\infty$, the dominant contribution is $\exp(-i
K H_{\mathrm{c}}t)$. One therefore considers the limiting evolution
operator
\begin{equation}
\label{eq:limevol} \cU(t)=\lim_{K\to\infty}\exp(i K
H_{\mathrm{c}}t)\,U_{K}(t),
\end{equation}
that can be shown to have the form
\begin{equation}
\label{eq:theorem} \cU(t)=\exp(-i H_{\mathrm{Z}} t),
\end{equation}
where
\begin{equation}
H_{\mathrm{Z}}=\hat P H =\sum_n P_n H P_n \label{eq:diagsys}
\end{equation}
is the Zeno Hamiltonian, $P_n$ being the eigenprojection of
$H_{\mathrm{c}}$ belonging to the eigenvalue $\eta_n$
\begin{equation}
\label{eq:diagevol}
H_{\mathrm{c}} = \sum_n \eta_n P_n, \qquad (\eta_n\neq\eta_m,
\quad \mbox{for} \; n\neq m) \ .
\end{equation}
This is formally identical to (\ref{eq:ZenoHam}) and
(\ref{eq:eqHz}). In conclusion, the limiting evolution operator is
\andy{eq:measinterbis}
\begin{eqnarray}
U_K(t)&\sim&\exp(-i K H_{\mathrm{c}}t)\,\cU(t)
\\
&=& \exp\left(-i\sum_n K  t\eta_n P_n + P_n H P_n t\right) ,
\label{eq:measinterbis}
\end{eqnarray}
whose block-diagonal structure is explicit. Compare with
(\ref{eq:evolNZ}). The above statements can be proved by making use
of the adiabatic theorem \cite{Messiah61}.

\section{A few comments on the three formulations}
\label{sec-comm}
\andy{comm}

The three different formulations of QZE summarized in the previous
sections are physically equivalent and the analogy among them can be
pushed very far \cite{FLP}.

After all, a projection \emph{\`a la} von Neumann is a handy way to
``summarize" the complicated physical processes that take place
during a quantum measurement. A measurement process is performed by
an external (macroscopic) apparatus and involves dissipative effects
(an interaction and an exchange of energy with and often a flow of
probability towards the environment). The external system performing
the observation need not be a \emph{bona fide} detection system,
namely a system that ``clicks" or is endowed with a pointer. For
instance, a spontaneous emission process is often a very effective
measurement process, for it is irreversible and leads to an
entanglement of the state of the system (the emitting atom or
molecule) with the state of the apparatus (the electromagnetic
field). The von Neumann rules arise when one traces away the
photonic state and is left with an incoherent superposition of
atomic states. However, it is clear that the main features of the
Zeno effects are still present if one formulates the measurement
process in more realistic terms, introducing a physical apparatus, a
Hamiltonian and a suitable interaction with the system undergoing
the measurement. It goes without saying that one can still make use
of projection operators, if such a description turns out to be
simpler and more economic (Occam's razor).

\begin{figure}[t]
\begin{center}
\includegraphics[height=6.5cm]{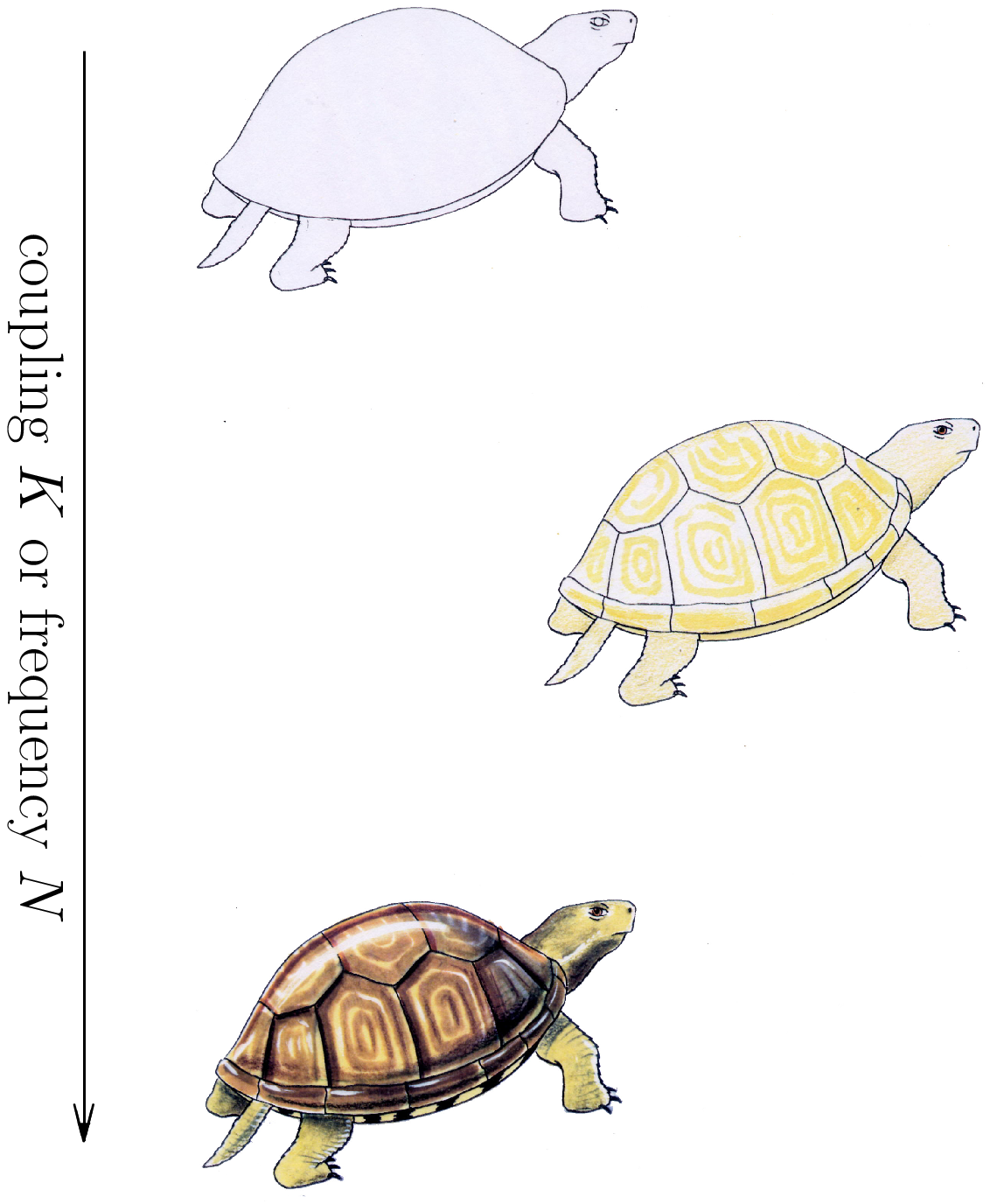}
\end{center}
\caption{\label{fig:tortoise} The Hilbert space of the system: a
dynamical superselection rule appears as the coupling $K$ to the
apparatus or the frequency $N$ of measurements is increased.}
\end{figure}

A comparison of the three Zeno procedures outlined in the three
preceding sections suggests that the continuous version might be the
most effective strategy against decoherence \cite{FTPNTL}. We shall
therefore look at some simple examples that yield a quantum Zeno
subspace by making use of a strong continuous coupling. As already
emphasized, the idea that a strong continuous coupling to an
external apparatus can yield a Zeno-like behavior has often been
proposed in the literature of the last two decades
\cite{varicont,zenoreview}.
The novelty in this context is the appearance of effective
superselection rules that are \textit{de facto} equivalent to the
celebrated ``W$^3$" ones \cite{WWW}, but turn out to be a mere
consequence of the Zeno dynamics. This remarkable aspect is
reminiscent of some results on ``classical" observables
\cite{Jauch}, the semiclassical limit \cite{Berry} and quantum
measurement theory \cite{Araki,Schwinger59}. The appearance of a
superselection rule in the Hilbert space of the system is
pictorially represented in Fig.\ \ref{fig:tortoise}. We conclude by
observing that the algebra of observables in the Zeno subspaces is
in itself an open problem and a topic worth investigating
\cite{FMPSS}.

\section{Example: three level system}
\label{sec-3lev}
\andy{3lev}

\begin{figure}
\begin{center}
\includegraphics[width=11cm]{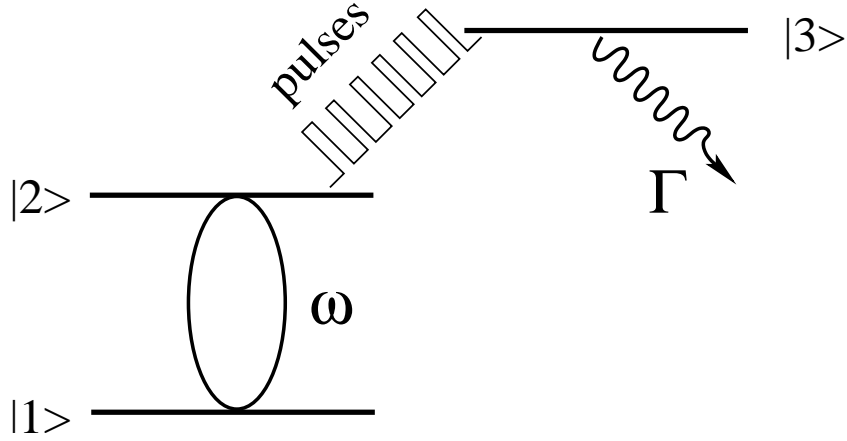}
\end{center}
\caption{``Pulsed" QZE, as in the experiment by Itano \emph{et al} \cite{Itano}.}
\label{fig:pulse}
\end{figure}

\begin{figure}[t]
\begin{center}
\includegraphics[width=11cm]{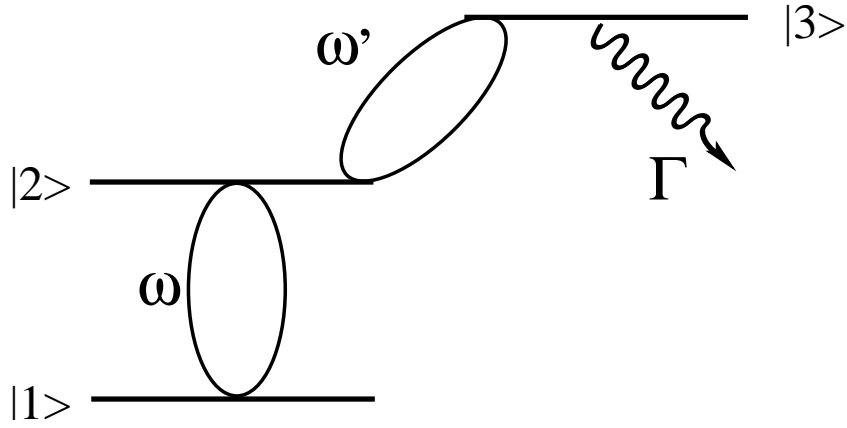}
\end{center}
\caption{``Continuous" observation, as in the experiment performed at MIT with a BEC \cite{ketterle}.}
\label{fig:cont}
\end{figure}

Let us now look at some simple examples, involving only
finite-dimensional Hilbert spaces. Let us first consider the
experiment schematically shown in Fig.\
\ref{fig:pulse}. Two levels 1 and 2 are Rabi coupled and frequent
pulses rapidly sweep the population from level 2 to level 3. If the
decay rate $\Gamma$ out of level 3 is fast enough, this can be
viewed as a very efficient measurement procedure and QZE takes
place. This is the scheme adopted by Itano and collaborators in
their celebrated experiment \cite{Itano}. Observe that in Fig.\
\ref{fig:pulse} one implicitly assumes $\pi/2$-pulses; on the other hand, if
one makes the pulses extremely frequent and accordingly less intense
(technically, one rescales the Hamiltonian \cite{FLP}), one gets a
situation similar to that shown in Fig.\ \ref{fig:cont}. In this
sense, both ``pulsed" and ``continuous" measurements are physically
equivalent in that they both yield QZE. This observation motivated
 a recent experiment performed at MIT with a BEC
\cite{ketterle}.

Consider now the scheme shown in Fig.\ \ref{fig:cont}, that
describes a three-level system undergoing two different Rabi
oscillations and coupled to the photon field. Photons drain the
population away from level 3 (towards some other level not shown in
the figure), so that if one restricts one's attention to the three
atomic levels 1, 2 and 3, the effective Hamiltonian in the Markovian
approximation reads $(\hbar = 1)$
\beq
H = \frac{\omega}{2} ( |1\rangle \langle 2| + |2\rangle \langle 1|)
+ \frac{\omega'}{2} ( |2\rangle \langle 3| + |3\rangle \langle 2|)
-i \frac{\Gamma}{2} |3\rangle \langle 3| = \frac{1}{2} \pmatrix{0 &
\omega & 0\cr \omega & 0 & \omega' \cr 0 & \omega' & -i\Gamma},
\label{eq:ham3leff}
\eeq
where
\beq
\langle 1| = (1,0,0), \quad
\langle 2| = (0,1,0), \quad
\langle 3| = (0,0,1).
\label{eq:3leveleff}
\eeq
This is the effective Hamiltonian that better describes the MIT
experiment  \cite{ketterle}. Once a $\Gamma$-photon has been
emitted, the atom recoils and leaves the condensate. Here and in the
following we will oversimplify the analysis in order to focus on the
main ideas.

The Hamiltonian (\ref{eq:ham3leff}) is not Hermitian and probability
is not conserved (the photon field has been traced out). Let us
prepare the system in state $|1\rangle$ and let
\beq
p_1(t) = |\langle 1| e^{-iH t} |1 \rangle |^2
\label{eq:surv3lev}
\eeq
be the survival probability. This is our key quantity, that can be
studied in different regimes.

\begin{figure}[t]
\begin{center}
\includegraphics[width=11cm]{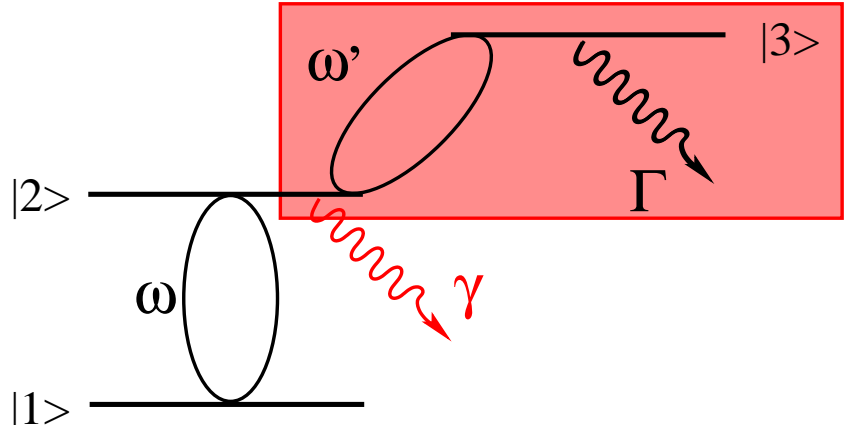}
\end{center}
\caption{Effective description of the system of Fig.\
\ref{fig:cont} in terms of only two levels: $\gamma = \omega'^{2}/\Gamma$.}
\label{fig:cont_eff}
\end{figure}

The main idea is that a larger coupling to the photon field (via
$\omega'$ and $\Gamma$) yields a more effective ``continuous"
observation of the population of level 2 (and therefore of level 1),
due to a quicker response of the environment/detection system. As
a consequence, the evolution is slower and one obtains QZE. We
assume \cite{ketterle}
\beq
\Gamma, \omega' \gg \omega,
\label{eq:ketterlevalues}
\eeq
which enables one to adopt the effective description of Fig.\
\ref{fig:cont_eff}: Consider only the two levels
\beq
\langle 1| = (1,0), \quad
\langle 2| = (0,1),
\label{eq:+-2}
\eeq
with the effective Hamiltonian
\beq
H_{{\rm eff}}= \frac{1}{2}\pmatrix{0 & \omega \cr \omega & -i\gamma}
=\frac{\omega}{2}(\ket{1}\bra{2}+\ket{2}\bra{1})-i\frac{\gamma}{2}
\ket{2}\bra{2},
\label{eq:nonhermham}
\eeq
with
\beq
\gamma = \omega'^{2}/\Gamma.
\eeq
This yields Rabi oscillations of frequency $\omega$, but at the same
time absorbs away the population of level $|2\rangle$, performing in
this way a ``measurement." Again, probabilities are not conserved.

An elementary calculation \cite{zenoreview} yields the survival
probability
\andy{survamplV}
\barr
p_{1}(t) = \left| \bra{1} e^{-i H_{{\rm eff}} t} \ket{1}\right|^2
&=&
\left| \frac{1}{2} \left( 1 + \frac{\gamma}{\sqrt{\gamma^2-4\omega^2}}
\right) e^{-(\gamma-\sqrt{\gamma^2-4\omega^2})t/4} \right.
\nonumber\\
& & + \left. \frac{1}{2} \left( 1 -
\frac{\gamma}{\sqrt{\gamma^2-4\omega^2}}
\right)e^{-(\gamma-\sqrt{\gamma^2+4\omega^2})t/4} \right|^2.
\label{eq:survamplV}
\earr
As expected, probability is (exponentially) absorbed away as $t
\to \infty$.
When $\gamma \gg \omega$ the above expression reads
\barr
p_1(t) = \left| \bra{1} e^{-i H_{\rm eff} t} \ket{1}\right|^2 \sim
e^{- t \omega^2/\gamma} = e^{-t \omega^2 \Gamma/\omega'^{2}} .
\label{eq:survamplV}
\earr
As expected, probability is (exponentially) absorbed away as $t \to
\infty$. However, as $\omega'$ (and hence $\gamma$) increases, the
effective decay rate $\omega^2/\gamma$ becomes smaller, eventually
halting the ``decay" (and consequent absorption) of the initial
state and yielding an interesting example of QZE: a larger $\gamma$
entails a more ``effective" measurement of the initial state.
Additional details on the derivation of the effective decay rate
$\omega^2/\gamma$ can be found in \cite{CT}. Notice that the
expansion (\ref{eq:survamplV}) is not valid at very short times
(where there is a quadratic Zeno region), but becomes valid very
quickly, on a time scale of order $\omega^{-1}$ (the duration of the
Zeno region
\cite{zenoreview,hydrogen}).

The (non-Hermitian) Hamiltonian (\ref{eq:nonhermham}) can be
obtained by considering the evolution engendered by a Hermitian
Hamiltonian acting on a larger Hilbert space and then restricting
the attention to the subspace spanned by $\{\ket{1}, \ket{2}\}$:
consider the Hamiltonian
\andy{hamflat}
\beq
\tilde H_\gamma= \Omega(\ket{1}\bra{2}+\ket{2}\bra{1}) +\int
d k\;k \ket{k} \bra{k} +\sqrt{\frac{\gamma}{2\pi}}\int dk\;(
\ket{2} \bra{k}+ \ket{k} \bra{2}) ,
\label{eq:hamflat}
\eeq
which describes a two-level system coupled to the photon field
$\{\ket{k}\}$ in the rotating-wave approximation. It is not
difficult to show \cite{zenoreview} that, if only state $\ket{1}$ is
initially populated, this Hamiltonian is equivalent to
(\ref{eq:nonhermham}), in that they both yield the \textit{same}
equations of motion in the subspace spanned by $\ket{1}$ and
$\ket{2}$. QZE is obtained by increasing $\gamma$: a larger coupling
to the environment leads to a more effective continuous observation
on the system (quicker response of the apparatus), and as a
consequence to a slower decay (QZE). The quantity $1/\gamma$ is the
response time of the apparatus.

\begin{figure}[t]
\begin{center}
\includegraphics[width=11cm]{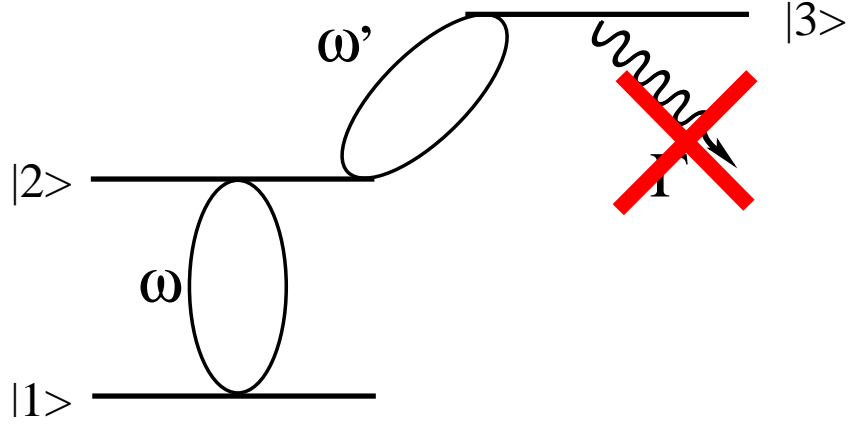}
\end{center}
\caption{Three stable levels, no dissipation.}
\label{fig:nodec3}
\end{figure}

We now discuss a different regime, yielding a different Zeno
picture. Rather than the situation (\ref{eq:ketterlevalues}),
consider
\beq
\Gamma \ll \omega'.
\label{eq:ourvalues}
\eeq
In this regime, the ``control" of the evolution, performed by
$\omega'$, is made with no dissipation/measurement. Let us first
discuss in which sense one gets a Zeno effect. Look at the idealized
case $\Gamma=0$ (Fig.\ \ref{fig:nodec3}). The (Hermitian)
Hamiltonian reads
\beq
H_{\mathrm{ideal}} = \frac{\omega}{2} ( |1\rangle \langle 2| +
|2\rangle \langle 1|) + \frac{\omega'}{2}  ( |2\rangle \langle 3| +
|3\rangle \langle 2|) = \frac{1}{2}\pmatrix{0 & \omega & 0\cr \omega
& 0 & \omega' \cr 0 & \omega' & 0} :
\label{eq:ham3l0}
\eeq
we shall argue that level 3 plays here the effective role of
measuring apparatus, although there is no \emph{bona fide}
measurement involved: as soon as the system is in $|2\rangle$ it
undergoes Rabi oscillations to $|3\rangle$. We expect level
$|3\rangle$  to perform better as a measuring apparatus when the
strength $\omega'$ of the coupling becomes larger.

\begin{figure}[t]
\begin{center}
\includegraphics[width=11cm]{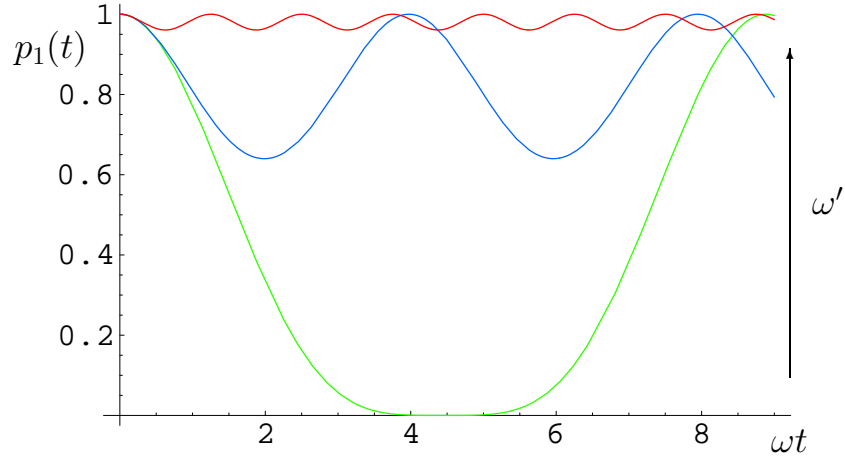}
\end{center}
\caption{Survival probability for a continuous Rabi coupling with
$\omega'=\omega$ (green), $\omega'=3 \,\omega$ (blue), $\omega'=9\,
\omega$ (red). }
\label{fig:zenocont}
\end{figure}

The survival probability in the initial state $|1\rangle$ reads
\beq
p_1(t) = \left| \bra{1} e^{-i H_{\mathrm{ideal}} t}
\ket{1}\right|^2= \frac{1}{(\omega'^2+\omega^2)^2}\left[\omega'^2+ \omega^2
\cos\left(\sqrt{\omega'^2+\omega^2}t/2\right) \right]^2 \label{eq:sp3}
\eeq
and is shown in Fig.\ \ref{fig:zenocont} for $\omega'=1,3,9\,
\omega$.
We notice that for large $\omega'$ the state of the system does not
change much: as $\omega'$ is increased, level $|3\rangle$ performs a
better ``observation" of the state of the system, hindering
transitions from $|1\rangle$ to $|2\rangle$. This can be viewed as a
QZE due to a continuous, Hermitian coupling to
level $|3\rangle$. This is the situation displayed in Fig.\
\ref{fig:nodec3}.

Note that one can also interpret the above results as an experiment
of electromagnetically induced transparency \cite{EIT}. Indeed, the
strong coupling  $\omega'$ between levels 2 and 3 splits state
$|2\rangle$ into two dressed states
$|\pm\rangle=(|2\rangle\pm|3\rangle)/\sqrt{2}$ of energy
$\pm\omega'$. For large values of $\omega'$ these states run out of
resonance and the system becomes transparent to the laser resonant
with the transition $1 \leftrightarrow 2$.

\begin{figure}[t]
\begin{center}
\includegraphics[width=11cm]{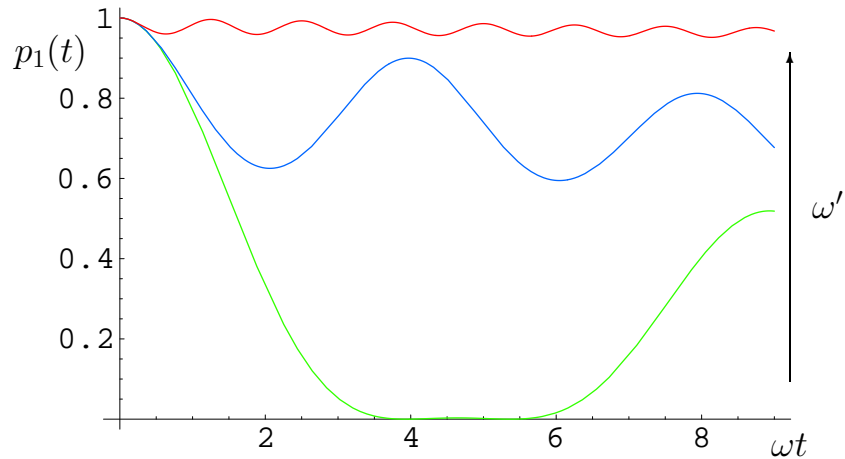}
\end{center}
\caption{Survival probability for a continuous Rabi coupling with a
small dissipation $\Gamma=0.2\,\omega$ on level \#3. Like before,
$\omega'=\omega$ (green), $\omega'=3\, \omega$ (blue), $\omega'=9\,
\omega$ (red). }
\label{fig:zenoalmost}
\end{figure}

But what happens for small $\Gamma$? An expansion yields
\beq
p_1(t) \stackrel{\Gamma \; {\rm small}}{\simeq}
\frac{1}{(\omega'^2+\omega^2)^2}\left[\omega'^2 e^{-\Gamma \omega^2
t/2(\omega^2+\omega'^2)} + \omega^2 e^{-\Gamma \omega'^2
t/4(\omega^2+\omega'^2)} \cos(\sqrt{\omega'^2+\omega^2}t/2)
\right]^2 .
\label{eq:sp3s}
\eeq
We stress that this is a very accurate formula (technically, the
expansion is uniform), essentially valid even for $\Gamma
=O(\omega)$ or $\Gamma =O(\omega')$ (or both!). The results are
shown Fig.\ \ref{fig:zenoalmost} for $\omega'=1,3,9 \,\omega$ and
$\Gamma=0.2\,\omega$.

In Figs.\ \ref{fig:zenocont} and \ref{fig:zenoalmost} we used
arbitrary values of the parameters for the sake of illustration. It
is remarkable how efficient the procedure turns out to be for more
physical values of $\omega, \omega', \Gamma$: one can get a
quantitative idea by plotting (with care!) Eq.\ (\ref{eq:sp3s}).

Clearly, in the very same regime (\ref{eq:ourvalues}), one can use a
pulsed technique, as in Fig.~\ref{fig:pulse}. This is sometimes
called ``bang-bang." In fact, it is nothing but pulsed Zeno.

\section{Some comments on interpretation}

\label{sec-applications}
\andy{applications}

Let us reinterpret the results of the previous section in the light
of the theorem proved in Sec.\ \ref{sec-contQZE}. Reconsider
Hamiltonian (\ref{eq:ham3l0})  (here $\omega'$ plays the role of $K$
in Sec.\ \ref{sec-contQZE})
\andy{ham3l}
\beq
H_{\mathrm{ideal}} = \frac{1}{2}\pmatrix{0 & \omega & 0\cr \omega &
0 &
\omega'
\cr 0 & \omega' & 0} = H+ \omega' H_{\mathrm{c}}, \label{eq:ham3l}
\eeq
where
\beq
H=\frac{\omega}{2} (\ket{1}\bra{2}+ \ket{2}\bra{1})=\frac{\omega}{2}
\pmatrix{
  0 & 1 & 0 \cr
  1 & 0 & 0 \cr
  0 & 0 & 0
},
\eeq
\beq
H_{\mathrm{c}}= \frac{1}{2}\ket{2}\bra{3}+\ket{3}\bra{2} =
\frac{1}{2}\pmatrix{
  0 & 0 & 0 \cr
  0 & 0 & 1 \cr
  0 & 1 & 0
}. \label{eq:Hmeas2}
\eeq
As $\omega'$ is increased, the Hilbert space is split into three
invariant subspaces (the eigenspaces of $H_{\mathrm{c}}$),
$\cH=\bigoplus\cH_{P_n}$, with
\beq
\cH_{P_0}=\mathrm{span}\{ \ket{1}\},\quad \cH_{P_1}= \mathrm{span}\{
\ket{2}+\ket{3}\}, \quad \cH_{P_{-1}}=\mathrm{span}\{
\ket{2}-\ket{3}\}, \label{eq:3sub}
\eeq
corresponding to the projections
\beq
P_0=\pmatrix{ 1 & 0 & 0 \cr 0 & 0 & 0 \cr 0 & 0 & 0 }, \quad
P_1=\frac{1}{2}\pmatrix{ 0 & 0 & 0 \cr 0 & 1 & 1 \cr 0 & 1 & 1 },
\quad P_{-1}=\frac{1}{2}\left(\begin{array}{rrr}
  0 & 0 & 0 \\
  0 & 1 & -1 \\
  0 & -1 & 1
\end{array}\right),
\eeq
with eigenvalues $\eta_0=0$ and $\eta_{\pm1}=\pm1$. The diagonal
part of the system Hamiltonian $H$ vanishes, $H_{\mathrm{diag}}=\sum
P_n H P_n=0$, and the Zeno evolution is governed by
\beq
\label{eq:HZ3lev}
H_{\mathrm{Z}}=\omega' H_{\mathrm{c}}=\frac{1}{2}\pmatrix{0 & 0 &
0\cr 0 & 0 & \omega' \cr 0 & \omega' & 0} .
\eeq
Any transition between $\ket{1}$ and $\ket{2}$ is inhibited: a
watched pot never boils.

\section{A watched cook can freely watch a boiling pot}
 \label{sec-cook}
 \andy{cook}

In the previous discussion we endeavored to make QZE on level 1.
Essentially, a Rabi transition from $\ket{1}$ to $\ket{2}$ was
inhibited by making use of different procedures. It would be
interesting if one could hinder transitions out of a Zeno
subspace, say a two-dimensional one, such as a qubit. Let us
therefore modify the level scheme, in order to understand whether
one can ``protect" the evolution of a qubit. Let
\beq
\langle 0| = (1,0,0,0), \quad
\langle 1| = (0,1,0,0), \quad
\langle 2| = (0,0,1,0), \quad
\langle 3| = (0,0,0,1)
\label{eq:4leveleff}
\eeq
and consider the four level Hamiltonian
\beq
H_{\mathrm{4lev}} =
\pmatrix{0 & \Omega & 0 & 0 \cr \Omega & 0 & \omega & 0 \cr 0 & \omega & 0 &
\omega' \cr 0 & 0 & \omega' & 0 }, \label{eq:ham4l}
\eeq
where states $\ket{0}$ and $\ket{1}$ undergo Rabi oscillations,
\beq
\Omega (\ket{0}\bra{1}+\ket{1}\bra{0})=
\pmatrix{0 & \Omega & 0 & 0 \cr \Omega & 0 & 0 & 0 \cr 0 & 0 & 0 & 0 \cr 0 &
0 & 0 & 0 } ,
\eeq
while state $\ket{2}$ ``observes" $\ket{1}$
\beq
\omega (\ket{1}\bra{2}+\ket{2}\bra{1})= \pmatrix{0 & 0& 0 &
0 \cr 0 & 0 & \omega  & 0 \cr 0 & \omega  & 0 & 0 \cr 0 & 0 & 0 & 0
}
\eeq
and state $\ket{3}$ ``observes"  $\ket{2}$
\beq
\omega' (\ket{2}\bra{3}+\ket{3}\bra{2})= \pmatrix{0 & 0&
0 & 0 \cr 0 & 0 & 0 & 0 \cr 0 & 0 & 0 & \omega' \cr 0 & 0 & \omega'
& 0 } .
\eeq
For the sake of simplicity, we assumed no dissipation and redefined
the Rabi frequencies in order to get rid of factors 2.

\begin{figure}[t]
\begin{center}
\includegraphics[width=11cm]{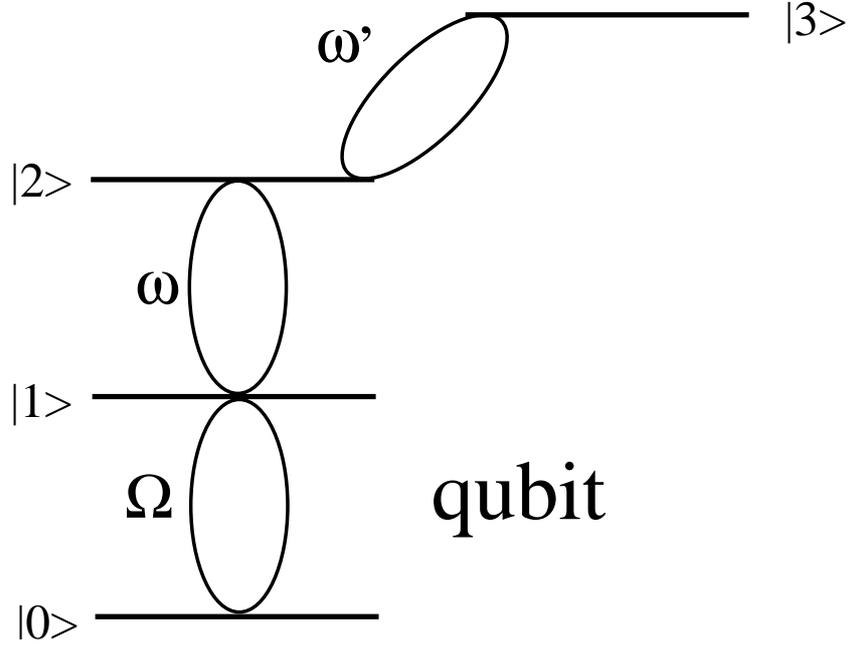}
\end{center}
\caption{``Control" of the dynamics of a qubit and suppression of
``decoherence."}
\label{fig:controlqubit}
\end{figure}

One can amuse oneself by deciding which ``observation" is more
effective \cite{theorem,Militello01}. In the following, it is
helpful to think of $\ket{0}$ and $\ket{1}$ as the states of a qubit
($\Omega$ being its dynamics), while $\omega$ is a caricature of a
``decoherence" process and $\omega'$ represents the ``control"
(whose objective is to suppress decoherence). See Fig.\
\ref{fig:controlqubit}.

Let us consider the worst situation, where $\omega \gg \Omega$ and
$\omega \gg \omega'$. Then one easily shows that the total Hilbert space $\cH$
splits into the three ``Zeno" subspaces:
\beq
\cH_0=\mathrm{span}\{ \ket{0}, \ket{3}\}, \quad \cH_+=\mathrm{span}\{
\ket{1}+\ket{2}\}, \quad \cH_-=\mathrm{span}\{
\ket{1}-\ket{2}\} \label{eq:3sub1}
\eeq
and the Zeno evolution is governed by
\beq
H_{\mathrm{4lev}} =
\pmatrix{0 & \Omega & 0 & 0 \cr \Omega & 0 & \omega & 0 \cr 0 & \omega & 0 &
\omega' \cr 0 & 0 & \omega' & 0 } \stackrel{\omega \; {\rm large}}{\longrightarrow}
H^{(\omega)}_{\mathrm{Z}}= \pmatrix{0 & 0 & 0 & 0
\cr 0 & 0 & \omega & 0 \cr 0 & \omega & 0 & 0 \cr 0 & 0 & 0 & 0 }.
\eeq
The Rabi oscillations between states $\ket{0}$ and $\ket{1}$ are
hindered (as well as those between $\ket{2}$ and $\ket{3}$). The
dynamics $\Omega$ of the qubit $\{\ket{0}, \ket{1}\}$ is destroyed
by decoherence $\omega$. This is detrimental.

On the other hand, if $\omega' \gg \omega$, and even if $\omega \gg
\Omega$, the total Hilbert space splits into the three
subspaces:
\beq
\cH'_{0}=\mathrm{span}\{ \ket{0}, \ket{1}\}, \quad \cH'_+=\mathrm{span}\{
\ket{2}+\ket{3}\}, \quad \cH'_-=\mathrm{span}\{
\ket{2}-\ket{3}\} \label{eq:3sub11}
\eeq
[notice the differences with (\ref{eq:3sub1})] and the Zeno
Hamiltonian reads
\beq
H_{\mathrm{4lev}} =
\pmatrix{0 & \Omega & 0 & 0 \cr \Omega & 0 & \omega & 0 \cr 0 & \omega & 0 &
\omega' \cr 0 & 0 & \omega' & 0 } \stackrel{\omega' \; {\rm large}}{\longrightarrow}
H^{(\omega')}_{\mathrm{Z}}=
\pmatrix{0 & \Omega & 0 & 0 \cr \Omega &
0 & 0 & 0 \cr 0 & 0 & 0 & \omega' \cr 0 & 0 & \omega' & 0 } .
\eeq
The conclusions are interesting and beautiful: the Rabi oscillations
between states $\ket{0}$ and $\ket{1}$ are fully \textit{preserved}
(\textit{even if and in spite of} $\omega \gg \Omega$). This is the
control of a qubit: its dynamics $\Omega$ is restored, even in
presence of strong decoherence $\omega >$ (or even $\gg$) $\Omega$.

In conclusion, if $\omega' \gg \omega, \Omega$, so that
$\omega'^{-1}$ is the shortest timescale in the problem,
\emph{a watched cook can freely watch a boiling pot.}
The suppression of the transition out of level 1 can be of great
help in protecting the dynamics of the qubit $\{\ket{0},
\ket{1}\}$. Unlike the one-dimensional case, an experiment on the
multidimensional QZE has a twofold objective: i) prevent probability
leakage \emph{out} of the Zeno subspace; ii) preserve the Zeno
dynamics \emph{in} the subspace. An experiment on the preservation
of the Zeno dynamics of a multidimensional system (such as a qubit)
has never been done.

\section{Conclusions}
\label{sec-concl}

We considered three physically equivalent formulations of the QZE
and focused on the ``continuous coupling" version, in order to
analyze some simple examples involving two, three and four level
systems. These examples are of interest in the light of possible
applications. Among these, we considered the possibility of
tailoring the interaction so as to obtain robust subspaces against
decoherence, useful for applications in quantum computation. The
experimental scheme proposed in Sec.\ \ref{sec-cook} would enable
one to control the dynamics of a qubit. It is remarkable that a
problem like the quantum Zeno effect, that was considered academic
until no more than 20 years ago, might lead in the next future to
novel experiments and very practical applications, motivating the
invention of techniques to counter decoherence and dissipation.

\ack
%\section*{Acknowledgments}
We would like to thank Wolfgang Ketterle for an interesting email
exchange. This work is partially supported by the European Union
through the Integrated Project EuroSQIP.

\end{document}